# Contact angles in the pseudopotential lattice Boltzmann modeling of wetting

Q. Li[1, 3], K. H. Luo[2, *], Q. J. Kang[3] and Q. Chen[4]

[1]Energy Technology Research Group, Faculty of Engineering and the Environment, University of Southampton, Southampton SO17 1BJ, United Kingdom

[2]Department of Mechanical Engineering, University College London, Torrington Place, London WC1E 7JE, United Kingdom

[3]Computational Earth Science Group, Earth and Environmental Sciences Division, Los Alamos National Laboratory, Los Alamos 87545, United States

[4]School of Energy and Power Engineering, Nanjing University of Science and Technology, Jiangsu 210094, China

[*]Corresponding author: K.Luo@ucl.uk

In this paper, we aim to investigate the implementation of contact angles in the pseudopotential lattice Boltzmann modeling of wetting at a large density ratio ($\rho_L/\rho_V = 500$). The pseudopotential lattice Boltzmann model [X. Shan and H. Chen, Phys. Rev. E **49**, 2941 (1994)] is a popular mesoscopic model for simulating multiphase flows and interfacial dynamics. In this model, the contact angle is usually realized by a fluid-solid interaction. Two widely used fluid-solid interactions: the density-based interaction and the pseudopotential-based interaction, as well as a modified pseudopotential-based interaction formulated in the present paper, are numerically investigated and compared in terms of the achievable contact angles, the maximum and the minimum densities, and the spurious currents. It is found that the *pseudopotential-based* interaction works well for simulating small static (liquid) contact angles ($\theta < 90°$), however, is unable to reproduce static contact angles close to $180°$. Meanwhile, it is found that the proposed *modified pseudopotential-based* interaction performs better in light of the maximum and the minimum densities and is overall more suitable for simulating large contact angles ($\theta > 90°$) as compared with the other two types of fluid-solid interactions. Furthermore, the spurious currents are found to be enlarged when the fluid-solid interaction force is introduced. Increasing the kinematic viscosity ratio between the vapor and liquid phases is shown to be capable of reducing the spurious currents caused by the fluid-solid interactions.

## I. Introduction

Wetting phenomena are widespread in natural and industrial processes [1-4]. The degree of wetting is determined by a force balance between the adhesive and cohesive forces. The adhesive force causes the liquid droplet to spread across the solid surface, while the cohesive force causes the droplet to ball up and avoid contact with the surface. The wettability of a solid surface by a liquid can be quantified via the contact angle. For smooth surfaces, the contact angle can be calculated from the Young's equation when the surface tensions are known. Small contact angles ($< 90°$) correspond to high wettability, whereas large contact angles ($> 90°$) correspond to low wettability [5].

Numerical studies of wetting phenomena involve the modeling of contact angles. In the conventional numerical approaches, the interaction between the flow and the solid wall is usually renormalized in terms of *ad hoc* boundary conditions for the hydrodynamic field [6]. For example, Ding and Spelt [7] have proposed a boundary condition named as geometric formulation for implementing contact angles in the phase-field method. For such a treatment, it is difficult to describe a variety of various solid properties with complex roughness landscape and chemical-physical attributes [6]. In the past two decades, the lattice Boltzmann (LB) method, which is a mesoscopic numerical approach based on the kinetic Boltzmann equation, has been developed into an alternative approach for simulating fluid flows and modeling physics of fluids [8-10]. Owing to its mesoscopic features, the LB method shows some distinctive advantages over the conventional numerical methods and has been widely used in modeling multiphase flows and interfacial phenomena [9, 10]. In the current multiphase LB community [9, 10], the pseudopotential LB model proposed by Shan and Chen [11] has attracted much attention because of its simplicity. In this model, the intermolecular interactions are represented with a density-dependent pseudopotential and the phase separation is achieved by imposing a short-range attraction between different phases.

Since the emergence of the pseudopotential LB model, there have been many studies of wetting phenomena using the model [12-18]. The first attempt was attributed to Martys and Chen in 1996 [12]. In their study, a fluid-solid interaction force was introduced to describe the interaction between the fluid and the solid wall. Different contact angles were obtained by adjusting the interaction strength of the fluid-solid interaction. Later, Raiskinmäki *et al*. [14, 15] proposed another type of fluid-solid interaction, which was reformulated by Sukop and Thorne in Ref. [8]. The pre-sum factor in the Martys and Chen's fluid-solid interaction is the density, while in the Raiskinmäki *et al*.'s fluid-solid interaction the pre-sum factor is the pseudopotential. These two types of fluid-solid interactions are therefore referred as the *density-based* interaction and the *pseudopotential-based* interaction, respectively.

In the literature, Kang *et al*. [16] have also formulated a fluid-solid interaction for the pseudopotential LB model. In addition, Benzi *et al*. [6] introduced a parameter, $\psi(\rho_{\mathrm{w}})$, which fixes the pseudopotential at the solid wall, to adjust the interaction between the fluid and the solid wall. In the next section, it will be shown that Kang *et al*.'s and Benzi *et al*.'s approaches can be classified into the above mentioned interactions. Recently, on the basis of the density-based interaction, Colosqui *et al*.

[18] proposed a new fluid-solid interaction, which is composed of a repulsive core and an attractive tail. Nevertheless, Colosqui *et al.* stressed that [18] their method is stable and robust within a range of moderate density ratios. The above-mentioned studies are related to the pseudopotential LB model. Here it is worth mentioning that in the LB community there are also a lot of studies in simulating wetting phenomena using other multiphase LB models, such as the free energy LB models [19-24].

In this paper, inspired by previous studies, we will formulate a *modified pseudopotential-based* fluid-solid interaction for the pseudopotential LB model. Subsequently, numerical simulations will be conducted to investigate the performance of the density-based interaction, the pseudopotential-based interaction, and the modified pseudopotential-based interaction. Specifically, the three types of fluid-solid interactions will be numerically compared in terms of the achievable contact angles, the maximum and the minimum densities, and the spurious currents. The rest of the present paper is organized as follows. In Section II, we will introduce the pseudopotential LB model as well as the fluid-solid interactions. Numerical results and analyses will be presented in Section III. Finally, a brief conclusion will be made in Section IV.

## II. Numerical model

### A. The MRT LB model

In this study, the LB model that uses a multi-relaxation-time collision operator [25] is employed. The time evolution of the MRT LB model can be written as follows:

$$f_\alpha \left( \boldsymbol{x} + \boldsymbol{e}_\alpha \delta_t, t + \delta_t \right) = f_\alpha \left( \boldsymbol{x}, t \right) - \left( \mathbf{M}^{-1} \mathbf{\Lambda} \mathbf{M} \right)_{\alpha\beta} \left( f_\beta - f_\beta^{eq} \right) + \delta_t F'_\alpha, \quad (1)$$

where $f_\alpha$ is the density distribution function, $f_\alpha^{eq}$ is its equilibrium distribution, $t$ is the time, $\boldsymbol{x}$ is the spatial position, $\boldsymbol{e}_\alpha$ is the discrete velocity along the $\alpha$ th direction, $\delta_t$ is the time step, $F'_\alpha$ represents the forcing term in the velocity space, $\mathbf{M}$ is an orthogonal transformation matrix, and $\mathbf{\Lambda}$ is a diagonal Matrix given by (for the D2Q9 lattice)

$$\mathbf{\Lambda} = \mathrm{diag}\left(\tau_\rho^{-1}, \tau_e^{-1}, \tau_\varsigma^{-1}, \tau_j^{-1}, \tau_q^{-1}, \tau_j^{-1}, \tau_q^{-1}, \tau_\nu^{-1}, \tau_\nu^{-1}\right). \tag{2}$$

Through the transformation matrix $\mathbf{M}$, the density distribution function $f_\alpha$ and its equilibrium distribution $f_\alpha^{eq}$ can be projected onto the moment space via $\mathbf{m} = \mathbf{M}\mathbf{f}$ and $\mathbf{m}^{eq} = \mathbf{M}\mathbf{f}^{eq}$, respectively. For the D2Q9 lattice, the equilibria $\mathbf{m}^{eq}$ can be given by

$$\mathbf{m}^{eq} = \rho\left(1, -2+3|\mathbf{v}|^2, 1-3|\mathbf{v}|^2, v_x, -v_x, v_y, -v_y, v_x^2-v_y^2, v_x v_y\right)^{\mathrm{T}}. \tag{3}$$

With Eqs. (2) and (3), the right-hand side of the MRT LB equation (1) can be rewritten as

$$\mathbf{m}^* = \mathbf{m} - \mathbf{\Lambda}\left(\mathbf{m} - \mathbf{m}^{eq}\right) + \delta_t\left(\mathbf{I} - \frac{\mathbf{\Lambda}}{2}\right)\mathbf{S}, \tag{4}$$

where $\mathbf{I}$ is the unit tensor and $\mathbf{S}$ is the forcing term in the moment space with $(\mathbf{I} - 0.5\mathbf{\Lambda})\mathbf{S} = \mathbf{M}\mathbf{F}'$.

The streaming process is given by

$$f_\alpha\left(\mathbf{x} + \mathbf{e}_\alpha \delta_t, t + \delta_t\right) = f_\alpha^*\left(\mathbf{x}, t\right), \tag{5}$$

where $\mathbf{f}^* = \mathbf{M}^{-1}\mathbf{m}^*$. The macroscopic density and velocity are calculated by

$$\rho = \sum_\alpha f_\alpha, \quad \rho\mathbf{v} = \sum_\alpha \mathbf{e}_\alpha f_\alpha + \frac{\delta_t}{2}\mathbf{F}, \tag{6}$$

where $\mathbf{F} = (F_x, F_y)$ is the total force acting on the system. The kinematic viscosity is given by $\upsilon = (\tau_\nu - 0.5)c_s^2 \delta_t$.

### B. The pseudopotential model and the fluid-solid interactions

The pseudopotential LB model was proposed by Shan and Chen [11]. In this model, the intermolecular interactions that cause phase segregation are mimicked by an intermolecular interaction force. The interaction force is defined via a pseudopotential $\psi$, which depends on the local fluid density. For single-component multiphase flows, the intermolecular interaction force is given by [26]

$$\mathbf{F}_m = -G\psi(\mathbf{x})\sum_\alpha w_\alpha \psi(\mathbf{x} + \mathbf{e}_\alpha)\mathbf{e}_\alpha, \tag{7}$$

where $G$ is the interaction strength and $w_\alpha$ are the weights. For the nearest-neighbor interactions on the D2Q9 lattice, the weights $w_\alpha = 1/3$ for $|e_\alpha|^2 = 1$ and $w_\alpha = 1/12$ for $|e_\alpha|^2 = 2$. According to Shan [21], Eq. (7) leads to the following pressure tensor [27]:

$$\mathbf{P} = \left[ \rho c_s^2 + \frac{Gc^2}{2}\psi^2 + \frac{Gc^4}{12}\psi\nabla^2\psi \right]\mathbf{I} + \frac{Gc^4}{6}\psi\nabla\nabla\psi , \qquad (8)$$

where $c$ is the lattice constant and $c_s = c/\sqrt{3}$ is the lattice sound speed. The last term on the right-hand side of Eq. (8) is related to the surface tension. Actually, the magic of the pseudopotential LB model lies in that the simple intermolecular interaction force is capable of not only modifying the equation of state ($p = \rho c_s^2 + Gc^2\psi^2/2$) but also yielding the surface tension.

The pseudopotential $\psi$ is taken as $\psi = \sqrt{2(p_{\text{EOS}} - \rho c_s^2)/Gc^2}$ [28], in which $p_{\text{EOS}}$ represents a prescribed equation of state. With this type of pseudopotential, the pseudopotential LB model usually suffers from thermodynamic inconsistency, i.e., the coexistence densities given by the pseudopotential LB model are inconsistent with the results given by the *Maxwell construction* [8]. A review of the pseudopotential LB model can be found in [29]. Recently, we found [27, 30] that the thermodynamic consistency can be approximately achieved in the pseudopotential LB model by adjusting the mechanical stability condition via an improved forcing scheme [27]:

$$\mathbf{S} = \begin{bmatrix} 0 \\ 6\mathbf{v}\cdot\mathbf{F} + \dfrac{12\sigma|\mathbf{F}_m|^2}{\psi^2 \delta_t (\tau_e - 0.5)} \\ -6\mathbf{v}\cdot\mathbf{F} - \dfrac{12\sigma|\mathbf{F}_m|^2}{\psi^2 \delta_t (\tau_\varsigma - 0.5)} \\ F_x \\ -F_x \\ F_y \\ -F_y \\ 2(v_x F_x - v_y F_y) \\ (v_x F_y + v_y F_x) \end{bmatrix}, \qquad (9)$$

where $|\mathbf{F}_m|^2 = (F_{m,x}^2 + F_{m,y}^2)$ and $\sigma$ is used to tune the mechanical stability condition. Note that $\mathbf{F}_m$ in Eq. (9) is the intermolecular interaction force while $\mathbf{F}$ is the total force of the system.

The intermolecular interaction force Eq. (7) represents the cohesive force of the system. When a solid wall is encountered, the adhesive force should also be considered. In 1996, Martys and Chen [12] introduced the following fluid-solid interaction to mimic the adhesive force in the pseudopotential LB model:

$$\mathbf{F}_{\text{ads}} = -G_{\text{w}} \rho(\mathbf{x}) \sum_{\alpha} \omega_{\alpha} s(\mathbf{x} + \mathbf{e}_{\alpha}) \mathbf{e}_{\alpha}, \tag{10}$$

where $G_{\text{w}}$ is the adsorption parameter, $\omega_{\alpha} = c_s^2 w_{\alpha}$, and $s(\mathbf{x} + \mathbf{e}_{\alpha})$ is a "switch" function, which is equal to 1 or 0 for a solid or a fluid phase, respectively. With the adhesive force $\mathbf{F}_{\text{ads}}$, the total force in Eq. (9) is given by $\mathbf{F} = \mathbf{F}_{\text{m}} + \mathbf{F}_{\text{ads}}$. Martys and Chen suggested that $G_w$ should be positive for non-wetting fluids and negative for wetting fluids. Different contact angles were obtained by adjusting $G_w$. Later, Raiskinmäki *et al*. [14, 15] proposed another type of fluid-solid interaction, which has been reformulated by Sukop and Thorne as follows [8]:

$$\mathbf{F}_{\text{ads}} = -G_{\text{w}} \psi(\mathbf{x}) \sum_{\alpha} \omega_{\alpha} s(\mathbf{x} + \mathbf{e}_{\alpha}) \mathbf{e}_{\alpha}. \tag{11}$$

It can be seen that the pre-sum factor in Eq. (10) is the density $\rho(\mathbf{x})$, while in Eq. (11) the pre-sum factor is the pseudopotential $\psi(\mathbf{x})$. As previously mentioned, these two types of fluid-solid interaction are therefore referred as the density-based interaction and the pseudopotential-based interaction, respectively.

Kang *et al*. [16] have also proposed a fluid-solid interaction for the pseudopotential LB model, which can be given as follows for single-component multiphase flows:

$$\mathbf{F}_{\text{ads}} = -G_{\text{w}} n(\mathbf{x}) \sum_{\alpha} n_{\text{w}}(\mathbf{x} + \mathbf{e}_{\alpha}) \mathbf{e}_{\alpha}. \tag{12}$$

In Ref. [16], $n(\mathbf{x})$ is the number density and $n_{\text{w}}(\mathbf{x} + \mathbf{e}_{\alpha})$ is a constant ($n_{\text{w}}$) at the wall and is zero elsewhere. Using the previous "switch" function $s(\mathbf{x} + \mathbf{e}_{\alpha})$, the "switch" function in Eq. (12) can be rewritten as $n_{\text{w}}(\mathbf{x} + \mathbf{e}_{\alpha}) = n_{\text{w}} s(\mathbf{x} + \mathbf{e}_{\alpha})$. In addition, Benzi *et al*.'s [6] have introduced a parameter $\psi(\rho_w)$ to fix the pseudopotential at the solid wall via a virtual wall density $\rho_w$, which can be summarized as follows [17]:

$$\mathbf{F}_{ads} = -G\psi(\mathbf{x})\sum_{\alpha}\omega_{\alpha}\psi(\rho_w)s(\mathbf{x}+\mathbf{e}_{\alpha})\mathbf{e}_{\alpha}, \qquad (13)$$

Since $n_w$ and $\rho_w$ are constants, $n_w$ and $\psi(\rho_w)$ can be actually absorbed into the constant $G_w$ or $G$, leading to an efficient constant $\hat{G}_w$. Hence Eqs. (12) and (13) can be classified into the density-based and the pseudopotential-based interactions, respectively.

Nevertheless, Eq. (13) gives us a hint: the "switch" function (when it is nonzero) can have the same order of magnitude as the pseudopotential. Meanwhile, $\psi(\rho_w)$ should not be a constant, otherwise, it can be absorbed into the constant $G_w$. In fact, Sukop and Thorne [8] have pointed out that the basic idea of the fluid-solid interaction is to create an analogue to the fluid-fluid interaction, Eq. (7). When the "switch" function is of the same order of magnitude as the pseudopotential $\psi$, the fluid-solid interaction may be more consistent with the fluid-fluid interaction. On the basis of the above consideration, the following modified pseudopotential-based interaction is proposed:

$$\mathbf{F}_{ads} = -G_w\psi(\mathbf{x})\sum_{\alpha}\omega_{\alpha}S(\mathbf{x}+\mathbf{e}_{\alpha})\mathbf{e}_{\alpha}, \qquad (14)$$

where the new "switch" function $S(\mathbf{x}+\mathbf{e}_{\alpha}) = \phi(\mathbf{x})s(\mathbf{x}+\mathbf{e}_{\alpha})$. The role of $\phi(\mathbf{x})$ is similar to that of $\psi(\rho_w)$ in Eq. (13). The difference is that $\psi(\rho_w)$ is a constant whereas $\phi(\mathbf{x})$ depends on $\mathbf{x}$. The simplest choice of $\phi(\mathbf{x})$ can be $\phi(\mathbf{x}) = \psi(\mathbf{x})$. With this choice, the "switch" function $S(\mathbf{x}+\mathbf{e}_{\alpha})$ is of the same order of magnitude as $\psi(\mathbf{x}+\mathbf{e}_{\alpha})$ in Eq. (7) when $s(\mathbf{x}+\mathbf{e}_{\alpha})$ is nonzero.

### III. Numerical results and discussion

In this section, numerical simulations are conducted to investigate the performance of the above three types of fluid-solid interactions. A piecewise linear equation of state [31] is adopted in the present study:

$$p(\rho) = \begin{cases} \rho\theta_V & \text{if } \rho \leq \rho_1 \\ \rho_1\theta_V + (\rho - \rho_1)\theta_M & \text{if } \rho_1 < \rho \leq \rho_2, \\ \rho_1\theta_V + (\rho_2 - \rho_1)\theta_M + (\rho - \rho_2)\theta_L & \text{if } \rho > \rho_2 \end{cases} \quad (15)$$

where $\theta_V = 0.64 c_s^2$, $\theta_L = c_s^2$, and $\theta_M = -0.04 c_s^2$. The unknown variables $\rho_1$ and $\rho_2$ in Eq. (15), which define the spinodal points, are obtained by solving a set of two equations: one for determining the mechanical equilibrium and the other for the chemical equilibrium [31]. The density ratio is set to 500 with $\rho_L = 500$ and $\rho_V = 1$, which corresponds to $\rho_1 = 1.36$ and $\rho_2 = 481.04$ [32]. The main requirement for $G$ in the pseudopotential $\psi = \sqrt{2(p_{EOS} - \rho c_s^2)/Gc^2}$ is to ensure that the whole term inside the square root is positive. Therefore we use $G = -1$ in this study. The parameter $\sigma$ in Eq. (9) is set to 0.084 in order to achieve thermodynamic consistency. The relaxation times are chosen as follows: $\tau_\rho = \tau_j = 1.0$, $\tau_e^{-1} = \tau_\varsigma^{-1} = 0.8$, and $\tau_q^{-1} = 1.1$. The lattice constant $c$ and the time step $\delta_t$ are both set to 1.

### A. The achievable contact angles

First, the three types of fluid-solid interactions are examined in terms of the achievable contact angles. In our simulations, a $N_x \times N_y = 300 \times 100$ lattice system is adopted. The periodical boundary conditions are employed in the y-direction while the non-slip boundary scheme [33] is applied at the solid wall: $f_5 = f_7 - 0.5(f_1 - f_3) - 0.25\delta_t(F_x + F_y)$, $f_6 = f_8 + 0.5(f_1 - f_3) + 0.25\delta_t(F_x - F_y)$, and $f_2 = f_4$. The density field is initialized as:

$$\rho(x,y) = \begin{cases} \rho_L & \text{if } \sqrt{(x-x_0)^2 + (y-y_0)^2} \leq R, \\ \rho_V & \text{else,} \end{cases} \quad (16)$$

where $(x_0, y_0) = (0.5N_x, 25)$ and $R = 30$. In simulations, the intermolecular interaction force given by Eq. (7) is also applied at the solid wall. For such a treatment, a contact angle of $\theta = 90°$ can be analytically obtained when setting $G_w = 0$, which represents the neutral wettability. The numerical solution given by $G_w = 0$ is shown in Fig. 1. Unless otherwise mentioned, the relaxation time $\tau_v$ is set to 1.1 for both the liquid and vapor phases (the kinematic viscosity $\upsilon = 0.2$). The obtained static contact angle is $\theta \approx 89.3°$, which is in good agreement with the analytical solution.

The static contact angles obtained by the $\rho$-based interaction, the $\psi$-based interaction, and the modified $\psi$-based interaction with different values of $G_w$ are shown in Figs. 2, 3, and 4, respectively. Form the figures we can see that all the three types of fluid-solid interactions are capable of modeling a wide range of contact angles through adjusting the parameter $G_w$. Meanwhile, it can be seen that the values of the parameter $G_w$ for the $\rho$-based interaction and the modified $\psi$-based interaction have the same order of magnitude, whereas the values of $G_w$ for the $\psi$-based interaction are an order of magnitude larger.

Furthermore, from Fig. 3 we can see that the $\psi$-based interaction gives a static contact angle $\theta = 165.6°$ when $G_w = 3.6$. It is expected that a larger contact angle can be achieved when $G_w$ increases. However, it is found that when $G_w \geq 3.65$ the $\psi$-based interaction cannot offer a static contact angle. To illustrate this point more clearly, we display the results of the cases $G_w = 3.6$ and $G_w = 3.65$ at $t = 1000\delta_t$, $3000\delta_t$, and $5000\delta_t$ in Fig. 5, from which it can be seen that in the case $G_w = 3.65$ the droplet will gradually detach from the solid wall. Similar phenomena are observed when $G_w > 3.65$. This means that under the considered condition the $\psi$-based interaction is unable to mimic static contact angles around $166°$ to $180°$. In contrast, the $\rho$-based interaction and the modified $\psi$-based interaction can successfully reproduce the contact angles close to $180°$. Their results for $\theta = 180°$ can be seen in Figs. 2(f) and 4(f), respectively. In fact, the capability of modeling large contact angles is of great importance in practical applications [34-36].

The results shown in Figs. 2, 3 and 4 are at the equilibrium state. Before reaching the equilibrium state, there is a dynamic spreading process after the liquid droplet contacts the wetting surface. In the literature, many studies [37-43] have revealed that the spreading (contact) radius usually follows a power-law scaling in time when the liquid droplet spreads over the surface. The Tanner's law ($r \sim t^{0.1}$ for 3D and $r \sim t^{1/7}$ for 2D when the surface tension at the contact line is the main driving force) [37-39], which describes the dynamics in the *final* stage of droplet spreading on a *completely* wetting surface ($\theta = 0$) [42, 43], is one of these power laws. For droplet spreading on a *partially* wetting surface, it has been found that [40-43] the dynamics in the *initial* stage also obeys the power law $r \sim t^n$, however, the exponent $n$ is non-universal and varies with the equilibrium contact angle. In the initial stage, the capillary driving force $\gamma(\cos\theta - \cos\theta^*)$ is very large, where $\gamma$ is the surface tension, $\theta$ is the equilibrium contact angle, and $\theta^*$ is the transient contact angle, hence it will lead to a high speed of spreading. Both experimental studies [40, 43] and molecular dynamics simulations [42] have shown that the exponent $n$ is around 0.5 when the equilibrium contact angle approaches zero and will decrease when the equilibrium contact angle increases.

By taking the modified $\psi-$based interaction as an example, here we test whether the dynamic wetting in our simulations obeys the power law. To eliminate the influence of the interface width, which is about four lattices in the present work, the droplet should be initially placed a finite distance away from the solid wall. The distance is equal to half of the interface width. Three different equilibrium contact angles are considered: $\theta = 19.0°$, $69.4°$, and $126.3°$ ($G_w = -0.3$, $-0.1$, and $0.2$, respectively). The dynamic wetting process at $\theta = 69.4°$ is displayed in Fig. 6. Quantitatively, the spreading (contact) radius is measured during the dynamic wetting process and the results are plotted in Fig. 7. Note that Fig. 7(b) is the log-log plot of the data given in Fig. 7(a). For comparison, the fitted curves $r/R = C(t/\tau_\rho)^n$ [40, 42, 43] are also shown in Fig. 7(b), where $\tau_\rho = \sqrt{\rho R^3/\gamma}$ is the characteristic inertial time. The considered regime is the inertial regime, in which the time $t$ is comparable to the characteristic inertial time $\tau_\rho$ and the spreading is dominated by inertia. From Fig. 7(b), we can see that the spreading radius basically follows the power law $r \sim t^n$ in all the considered cases. The fitted exponents for the cases $\theta = 19.0°$, $69.4°$, and $126.3°$ are given by $n = 0.51$, $0.485$, and $0.425$, respectively. Clearly, our results also show that the exponent $n$ decreases with the increase of the equilibrium contact angle and is around 0.5 when the equilibrium contact angle approaches zero.

### B. The maximum and minimum densities

In the pseudopotential LB model, the coexistence liquid and vapor densities are mainly determined by the mechanical stability condition of the model [26]. When the fluid-solid interaction force $\mathbf{F}_{ads}$ is introduced and the parameter $G_w$ is nonzero, the mechanical stability condition near the solid wall will be changed. Correspondingly, the coexistence densities will also be changed. As a result, the minimum and maximum densities of the system will deviate from the equilibrium vapor and liquid densities, respectively.

The minimum and maximum densities obtained by the three types of fluid-solid interactions for the cases $\theta \approx 30°$, $60°$, $90°$, $120°$, $150°$, and $180°$ are listed in Table I. From the table we can see that in the case $\theta \approx 90°$ ($G_w \approx 0$) the minimum and maximum densities are basically equal to the corresponding equilibrium vapor and liquid densities ($\rho_V = 1$ and $\rho_L = 500$), respectively. For the cases $\theta > 90°$ ($G_w > 0$) we can see that the maximum densities $\rho_{max}$ agree well with the liquid density $\rho_L$, however, the minimum densities $\rho_{min}$ gradually deviate from the vapor density $\rho_V$ when the contact angle increases. On the contrary, when $\theta < 90°$, the maximum densities will deviate from $\rho_L$, with the minimum densities being in good agreement with $\rho_V$.

Furthermore, we can see that, for different types of fluid-solid interactions, the deviations are different. When $\theta < 90°$, the $\psi-$based interaction gives relatively small deviations. However, when $\theta > 90°$, the minimum density $\rho_{\min}$ given by the $\psi-$based interaction significantly deviates from the equilibrium vapor density $\rho_V$ when the contact angle increases. Meanwhile, we can see that, for the cases $\theta > 90°$, the modified $\psi-$based interaction performs much better than the $\psi-$based interaction and the $\rho-$based interaction in terms of the minimum densities. For example, in the case $\theta \approx 150°$ the relative deviations (between $\rho_{\min}$ and $\rho_V$) given by the modified $\psi-$based, the $\rho-$based, and the $\psi-$based interactions are about $4\%$, $13\%$, and $79\%$, respectively. The different performance of the three types of fluid-solid interactions may be due to the fact that the mechanical stability condition is changed differently. In addition, from Table I we can find that the minimum density $\rho_{\min}$ given by the $\psi-$based interaction will approach zero when the contact angle increases. This may be the reason why the $\psi-$based interaction is unable to mimic the contact angles close to $180°$.

For completeness, we also compare the performance of the $\rho-$based and the modified $\psi-$based interactions on rough surfaces in terms of the minimum and maximum densities. The used computational system is $Nx \times Ny = 300 \times 150$. The rough surface is generated with square pillars. The height of the pillars is 20 lattices, the width of the pillars is 6 lattices, and the spacing between pillars is 3 lattices. Initially, a droplet with $R = 30$ is placed on the rough surface. Note that, when simulations on rough surfaces are involved, it is not recommended to apply the intermolecular interaction force Eq. (7) at solid walls since it is inconvenient to implement ghost layers below complex rough surfaces. Usually, for a given contact angle on flat surfaces, the values of $G_w$ are different between the cases with and without the intermolecular interaction force at solid walls.

The results of the modified $\psi$–based interaction ($G_w = 0.375$) and the $\rho$–based interaction ($G_w = 0.255$) on the prescribed rough surface are shown in Fig. 8. Their corresponding static contact angles on flat surfaces are about $138.5°$ and $138°$, respectively. From Fig. 8 we find that the contact angles on the rough surface given by the both interactions are approximately equal to $146°$. According to the Cassie's law [44-46], the apparent contact angle of a droplet that sits on the roughness peaks is given by

$$\cos\theta_C = f'\cos\theta + f' - 1, \tag{17}$$

where $\theta$ is the equilibrium contact angle on flat surfaces and $f'$ is the fraction of solid surface area on the horizontal projected plane. In three-dimensional space, $f'$ is defined as $f' = a^2/(a+b)^2$ for rough surfaces with square pillars [45, 46], where $a$ is the width of pillars and $b$ is the spacing between pillars. For two-dimensional simulations, it is obvious that $f'$ should be calculated by $f' = a/(a+b)$. In the present work $a = 6$ and $b = 3$, hence $f' = a/(a+b) = 6/9 \approx 0.6667$. According to Eq. (17), for $\theta \approx 138.5°$ and $138°$, $f' = 6/9$ gives $\theta_C \approx 146.4°$ and $\theta_C \approx 146.0°$, respectively. It is seen that our numerical results agree well with the prediction of the Cassie's law.

The detailed comparisons of the $\rho$–based and the modified $\psi$–based interactions are made in Table II. From the table we can see that, in the considered case, the maximum densities $\rho_{max}$ given by the $\rho$–based and the modified $\psi$–based interactions are both in good agreement with the equilibrium liquid density $\rho_L$, while their minimum densities deviate from the equilibrium vapor density $\rho_V$. Nevertheless, the deviations are different for the two interactions. To be specific, the deviations (between $\rho_{min}$ and $\rho_V$) on the rough surface are about 18% and 6.1% for the $\rho$–based and the modified $\psi$–based interactions, respectively. Meanwhile, we can see that there is a considerable change in $\rho_{min}$ for the $\rho$–based interaction between the results on flat and rough surfaces, but there is just a slight change for the modified $\psi$ – based interaction. In summary, it is found that the modified $\psi$ – based interaction also performs better on rough surfaces in terms of the minimum and maximum densities.

### C. The spurious currents

The so-called spurious currents, which are also known as parasitic velocities in the vicinity of interfaces, exist in almost all the numerical simulations of multiphase flows and interfacial phenomena. In this section, we investigate the spurious currents given by the three types of fluid-solid interactions. The maximum magnitudes of the spurious currents produced by the $\rho$ – based, the $\psi$ – based, and the modified $\psi$ – based interactions are shown in Fig. 9.

The values of the parameter $G_w$ are taken in the intervals $[-0.14, 0.28]$, $[-2, 3]$, and $[-0.2, 0.4]$ for the $\rho$ – based, the $\psi$ – based, and the modified $\psi$ – based interactions, respectively. Within the above intervals, the smallest contact angles given by the three types of fluid-solid interactions are around $45°$ [see Figs. 2(b), 3(b), and 4(b)] and the largest contact angles are about $150° \sim 160°$. According to Fig. 9, the following information can be obtained. First, it can be seen that the maximum magnitudes of the spurious currents significantly increase when the relaxation time $\tau_v$ decreases, which is expected since such a feature has been found in previous studies without solid walls [47]. Furthermore, we can see that, with the increase of $|G_w|$, the maximum magnitudes of the spurious currents will increase for all three types of interactions and it can be seen that $G_w = 0$ gives the smallest values, which indicates that the spurious currents of the pseudopotential LB model are enlarged when the fluid-solid interaction force is included.

Moreover, by comparing Fig. 9(b) with Figs. 9(a) and 9(c), we can see that, although the $\psi$ – based interaction also gives large spurious currents when $G_w > 0$, it yields relatively small spurious currents when $G_w < 0$, in comparison with the results given by the $\rho$ – based and the modified $\psi$ – based interactions. According to the present results and the results shown in the previous sections, we can conclude that the $\psi$ – based interaction is more suited for simulating small static contact angles $\theta < 90°$.

Meanwhile, from Figs. 9(a) and 9(c) we can find the maximum magnitudes of the spurious currents given by the $\rho$ – based and the modified $\psi$ – based interactions vary similarly against the parameter $G_w$. Actually, their results are nearly the same for the same contact angle. In this sense, the *$\rho$ – based interaction* can be replaced by the *modified $\psi$ – based interaction* since they have almost the same performance according to the achievable contact angles as well as the spurious currents, and the modified $\psi$ – based interaction gives better results in light of the maximum and minimum densities. On the basis of the numerical results, we believe that the modified $\psi$ – based interaction is overall more suitable for simulating contact angles $\theta > 90°$ as compared with the other two types of fluid-solid interactions.

Finally, we would like to discuss the reduction of the spurious currents caused by the fluid-solid interactions. Obviously, when the spurious currents are as large as the characteristic velocities of the simulated problem, the ambiguity between the physical and spurious velocities will appear. Previously, the relaxation time $\tau_v$ is set to the same value for both the liquid and vapor phases, which gives $\upsilon_V/\upsilon_L = 1$ ($\upsilon_V/\upsilon_L$ is the kinematic viscosity ratio between the vapor and liquid phases). Under such a condition, the dynamic viscosity ratio $\mu_L/\mu_V = (\rho_L/\rho_V)/(\upsilon_V/\upsilon_L)$ is equal to the density ratio $\rho_L/\rho_V$. Accordingly, a large density ratio will yield a large dynamic viscosity ratio, which will lead to a sharp change of the viscous stress tensor near the interface [27].

Here we show that increasing the kinematic viscosity ratio $\upsilon_V/\upsilon_L$ (also the decrease of the dynamic viscosity ratio $\mu_L/\mu_V$) can reduce the spurious currents caused by the fluid-solid interactions. For simplicity, we only take the modified $\psi$–based interaction as an example. The liquid kinematic viscosity is fixed at $\upsilon_L = 1/30$ (corresponding to $\tau_v = 0.6$ in the above simulations). The maximum magnitudes of the spurious currents given by three different values of $\upsilon_V/\upsilon_L$ (1, 5, and 10) are shown in Fig. 10. From the figure we can see that the maximum magnitudes of the spurious currents can be gradually reduced with the increase of $\upsilon_V/\upsilon_L$. To show the reduction more clearly, the velocity vectors of the case $G_w = 0.4$ with $\upsilon_V/\upsilon_L = 1$ and 10 are displayed in Fig. 11 together with the corresponding interfaces. From the figure we can see that, when $\upsilon_V/\upsilon_L$ increases from 1 to 10, the spurious currents are noticeably reduced. In addition, from Fig. 11 we can find that in the liquid phase the spurious currents are much smaller than those in the vapor phase. To be specific, in the case $G_w = 0.4$ with $\upsilon_V/\upsilon_L = 10$ the maximum magnitude of the spurious currents in the liquid phase is about 0.012.

## IV. Conclusions

In this paper, we have studied the implementation of contact angles in the pseudopotential LB modeling of wetting at a large density ratio ($\rho_L/\rho_V = 500$). A modified pseudopotential-based fluid-solid interaction has been formulated for the pseudopotential LB model. Two previous fluid-solid interactions and the proposed modified pseudopotential-based interaction have been investigated and compared. The main findings and conclusions are summarized as follows.

(1): In terms of the achievable contact angles, the $\rho$–based interaction and the modified $\psi$–based interaction are better than the $\psi$–based interaction.

(2): The minimum density $\rho_{\min}$ given by the $\psi$–based interaction is found to approach zero when the contact angle increases, which may be the reason why the $\psi$–based interaction is unable to mimic static contact angles close to $180°$.

(3): In light of the maximum and minimum densities, the modified $\psi$–based interaction performs better than the $\rho$–based interaction on both flat and rough surfaces.

(4): For all the three types of fluid-solid interactions, the maximum magnitudes of the spurious currents are found to increase with the increase of $|G_w|$, which means that the spurious currents of the pseudopotential LB model are enlarged when the fluid-solid interaction force is included.

(5): It is found that the spurious currents caused by the fluid-solid interaction force can be considerably reduced via increasing the kinematic viscosity ratio between the vapor and liquid phases.

According to the results, when the pseudopotential LB model is applied to simulate (liquid) wetting phenomena at large density ratios, we suggest using the $\psi$–based interaction for simulating small static contact angles. For large static contact angles, the modified $\psi$–based interaction is overall the most suitable one among the three types of fluid-solid interactions. Furthermore, increasing the kinematic viscosity ratio between the vapor and liquid phases can suppress the spurious currents caused by the fluid-solid interactions. Finally, we would like to point out that investigating the maximum and minimum densities as well as the spurious currents are applicable to all the treatments that are devised to implement contact angles in the pseudopotential LB modeling.

## Acknowledgments

The authors gratefully acknowledge the support from the Engineering and Physical Sciences Research Council of the United Kingdom under Grant Nos. EP/I012605/1 and EP/J016381/2, and the Los Alamos National Laboratory's Lab Directed Research & Development (LDRD) Program.

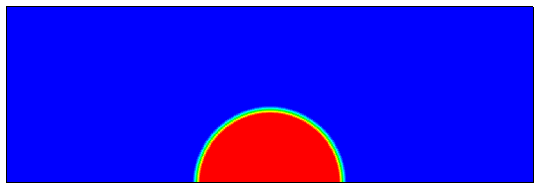

FIG. 1. (Color online) The contact angle at $G_w = 0$.

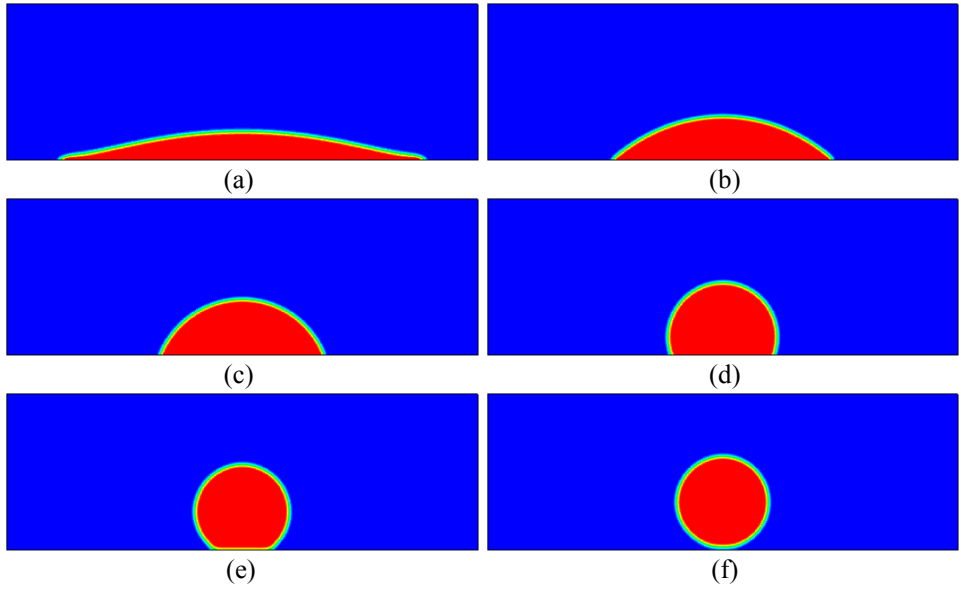

FIG. 2. (Color online) Contact angles obtained with the $\rho-$ based interaction: (a) $G_w = -0.21$, $\theta = 17.6°$, (b) $G_w = -0.14$, $\theta = 43.3°$, (c) $G_w = -0.07$, $\theta = 68.6°$, (d) $G_w = 0.07$, $\theta = 108.7°$, (e) $G_w = 0.21$, $\theta = 144.6°$, and (f) $G_w = 0.36$, $\theta = 180°$.

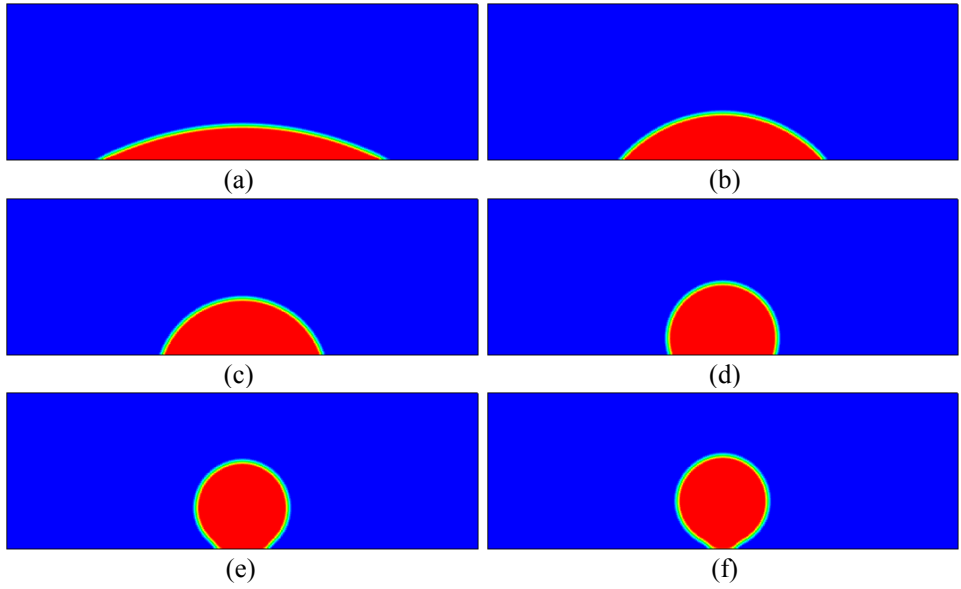

FIG. 3. (Color online) Contact angles obtained with the $\psi-$based interaction: (a) $G_w = -3.0$, $\theta = 27.2°$, (b) $G_w = -2.0$, $\theta = 49.6°$, (c) $G_w = -1.0$, $\theta = 70.6°$, (d) $G_w = 1.0$, $\theta = 107.5°$, (e) $G_w = 3.0$, $\theta = 147.9°$, and (f) $G_w = 3.6$, $\theta = 165.6°$.

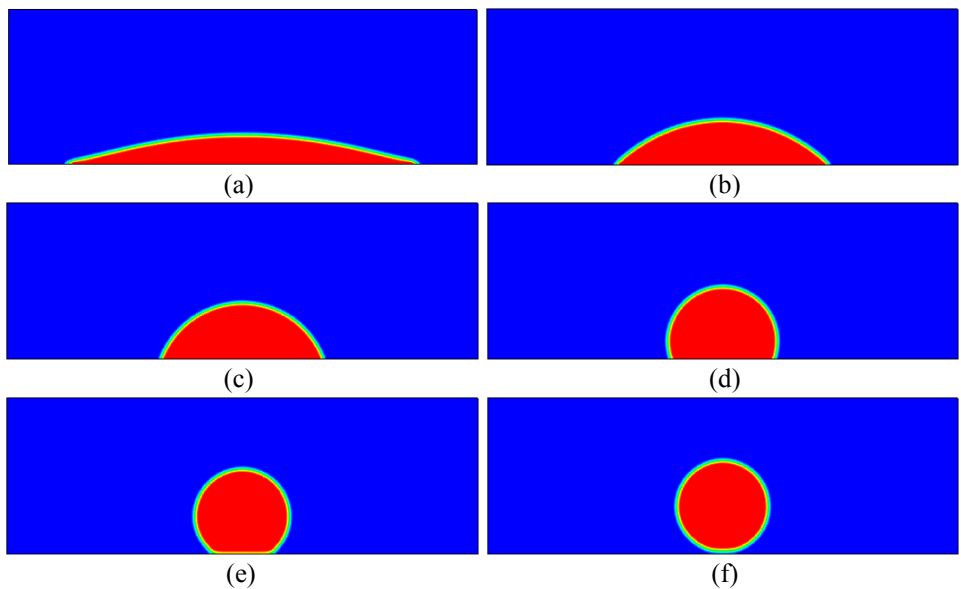

FIG. 4. (Color online) Contact angles obtained with the modified $\psi-$ based interaction: (a) $G_w = -0.3$, $\theta = 19.0°$, (b) $G_w = -0.2$, $\theta = 45.9°$, (c) $G_w = -0.1$, $\theta = 69.4°$, (d) $G_w = 0.1$, $\theta = 108.3°$, (e) $G_w = 0.3$, $\theta = 143.1°$, and (f) $G_w = 0.53$, $\theta = 180°$.

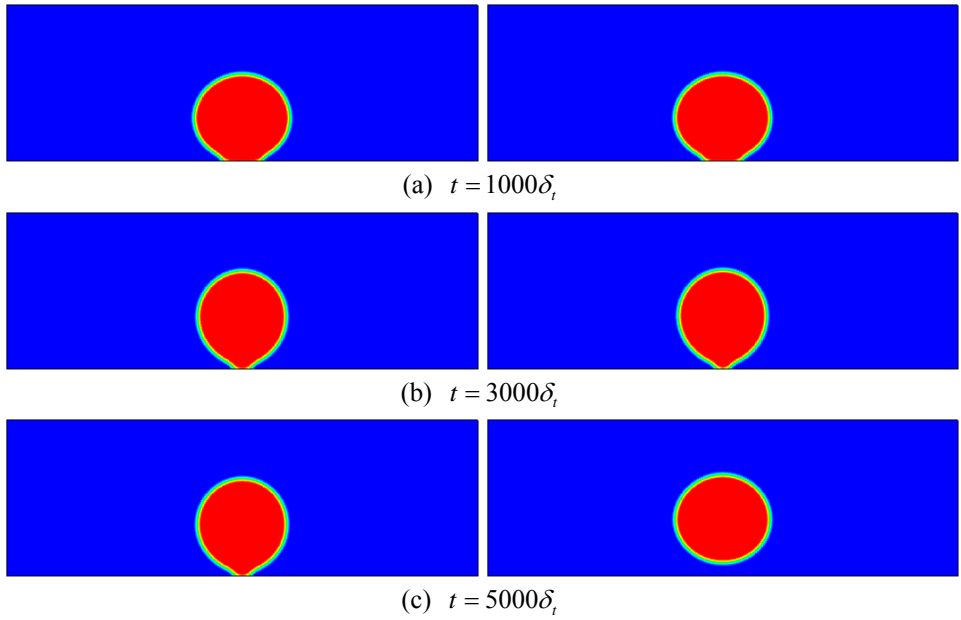

(a) $t = 1000\delta_t$

(b) $t = 3000\delta_t$

(c) $t = 5000\delta_t$

FIG. 5 (Color online) The results of the $\psi$ – based interaction with $G_w = 3.6$ (left) and $G_w = 3.65$ (right) at $t = 1000\delta_t$, $3000\delta_t$, and $5000\delta_t$.

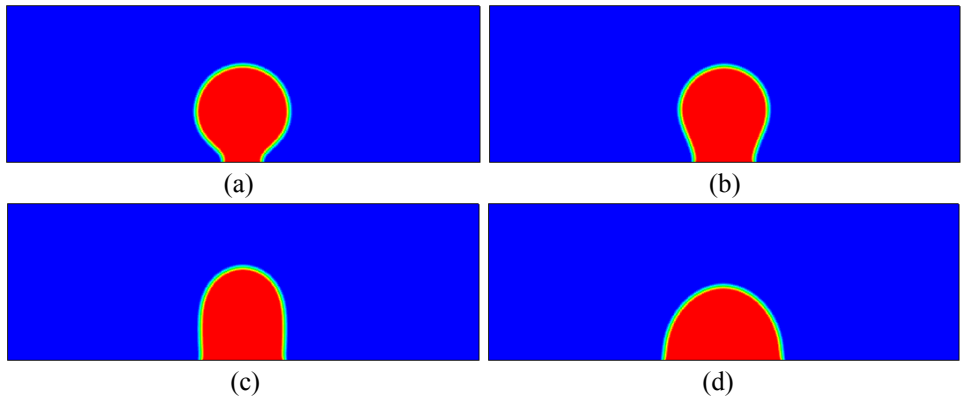

FIG. 6. (Color online) Dynamic wetting process at $\theta = 69.4°$: (a) $t = 200\delta_t$, (b) $t = 500\delta_t$, (c) $t = 1000\delta_t$, and (d) $t = 2000\delta_t$.

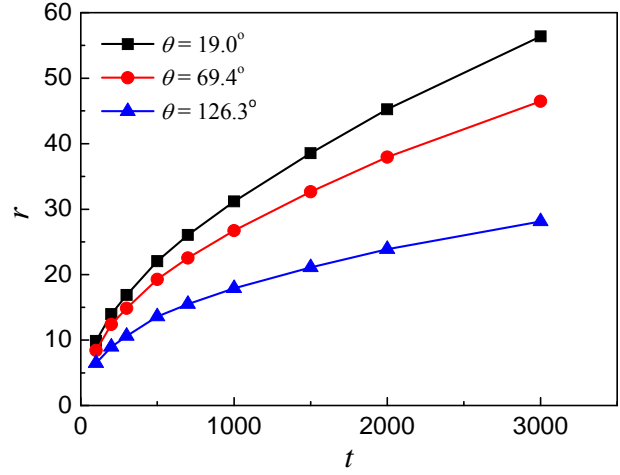

(a)

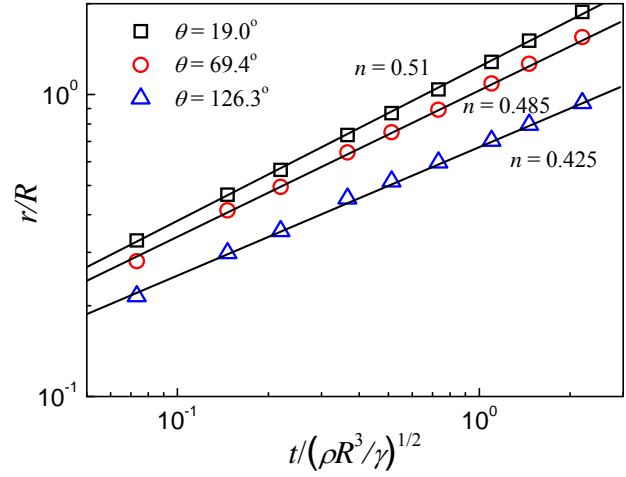

(b)

FIG. 7. (Color online) The spreading (contact) radius measured as a function of time for three different equilibrium contact angles: (a) standard plot and (b) log-log plot together with the fitted curves $r/R = C\left(t/\tau_\rho\right)^n$, where $\tau_\rho = \sqrt{\rho R^3/\gamma}$.

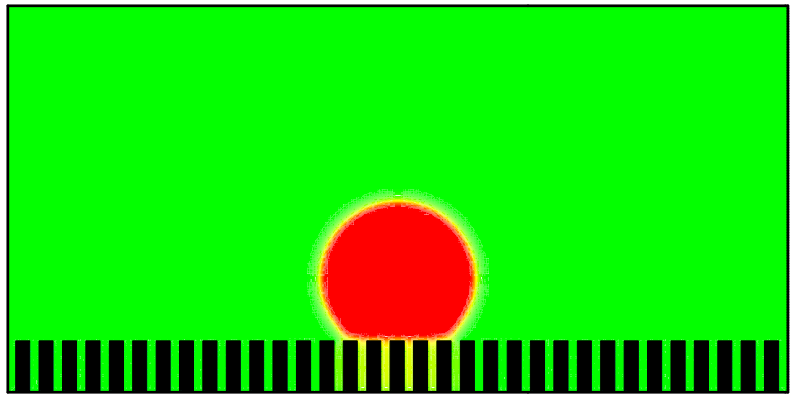

(a) the modified $\psi$ – based

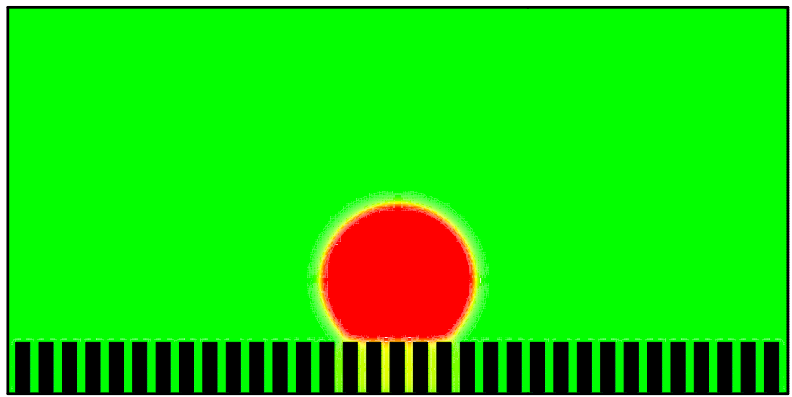

(b) the $\rho$ – based

FIG. 8 (Color online) The results of the modified $\psi$ – based and the $\rho$ – based interactions on rough surfaces. The liquid phase, vapor phase, and the solid surfaces are denoted by red, light green, and black, respectively. For both interactions, their corresponding equilibrium contact angles on flat surface are around $\theta = 138°$.

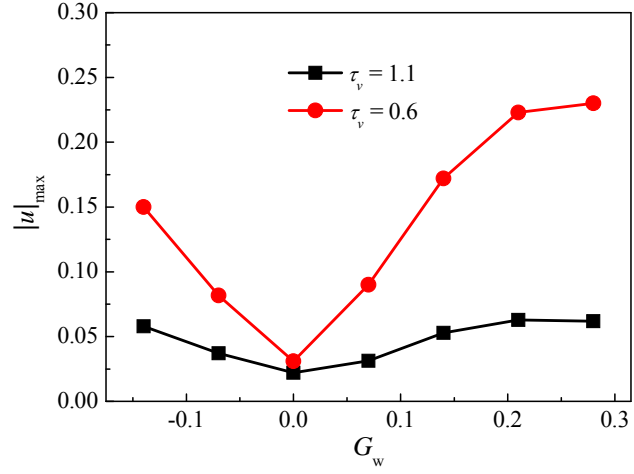

(a)

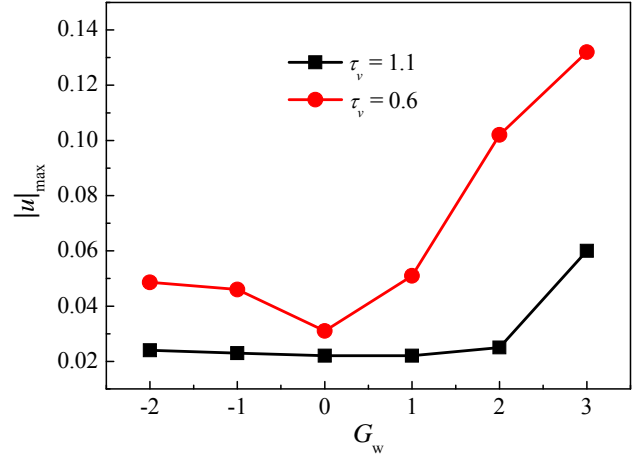

(b)

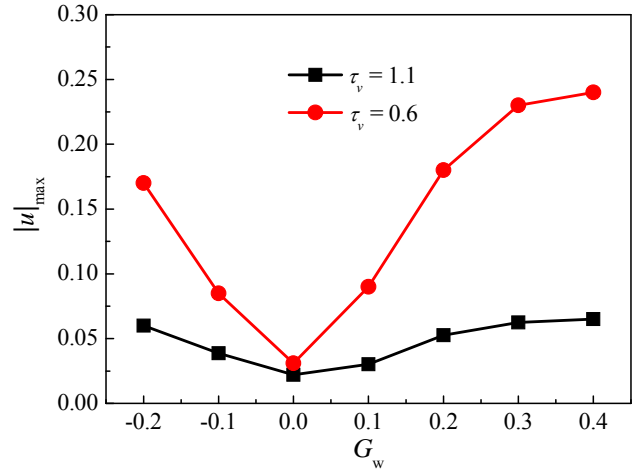

(c)

FIG. 9. (Color online) The maximum magnitudes of the spurious currents obtained with different fluid-solid interactions: (a) the $\rho$ – based interaction, (b) the $\psi$ – based interaction, and (c) the modified $\psi$ – based interaction.

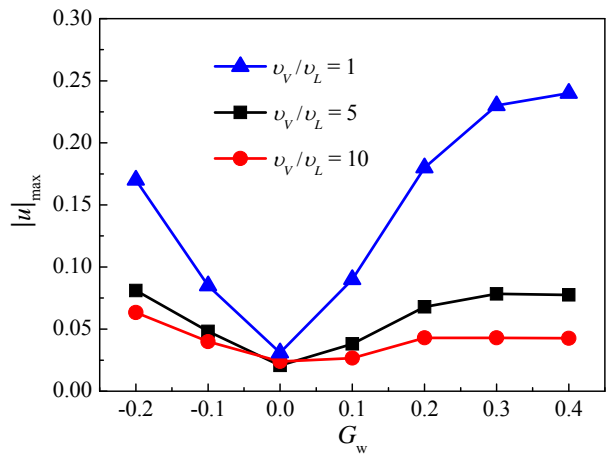

FIG. 10. (Color online) The maximum magnitudes of the spurious currents obtained by the modified $\psi$ − based interaction with different kinematic viscosity ratios $\upsilon_V/\upsilon_L$.

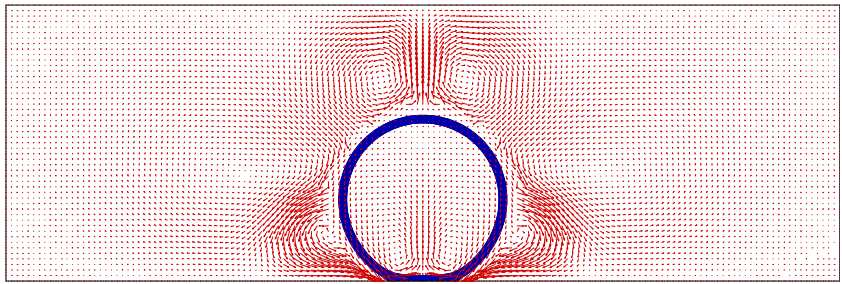

(a)

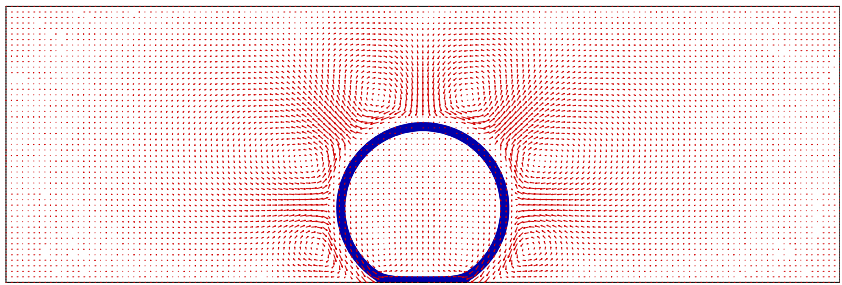

(b)

FIG. 11. (Color online) The velocity vectors obtained with the modified $\psi-$based interaction at $G_\mathrm{w} = 0.4$. (a) $\upsilon_V/\upsilon_L = 1$ and (b) $\upsilon_V/\upsilon_L = 10$.

**Table I** Comparisons of the minimum and maximum densities between different types of interactions.

| $\theta \approx$ (degrees) | $\rho$ – based | | $\psi$ – based | | modified $\psi$ – based | |
|:---:|:---:|:---:|:---:|:---:|:---:|:---:|
| | $\rho_{min}$ | $\rho_{max}$ | $\rho_{min}$ | $\rho_{max}$ | $\rho_{min}$ | $\rho_{max}$ |
| 30 | 0.99 | 545.9 | 0.98 | 526.6 | 1.00 | 543.4 |
| 60 | 1.00 | 524.2 | 1.00 | 514.0 | 1.00 | 523.7 |
| 90 | 1.00 | 500.5 | 1.00 | 500.5 | 1.00 | 500.5 |
| 120 | 0.94 | 501.0 | 0.52 | 500.7 | 0.98 | 501.0 |
| 150 | 0.87 | 501.2 | 0.21 | 500.8 | 0.96 | 501.2 |
| 180 | 0.79 | 501.7 | — | — | 0.93 | 501.7 |

**Table II** Comparisons of the results between the modified $\psi$ – based and $\rho$ – based interactions.

| interaction | $G_w$ | rough surface (flat surface) | | |
|:---:|:---:|:---:|:---:|:---:|
| | | $\rho_{min}$ | $\rho_{max}$ | contact angle |
| $\rho$ – based | 0.255 | 0.820 (0.853) | 500.9 (500.8) | 146 (138) |
| modified $\psi$ – based | 0.375 | 0.939 (0.942) | 500.9 (500.8) | 146 (138.5) |